\documentclass[prd,twocolumn,showpacs,preprintnumbers,nofootinbib,amsmath,amssymb]{revtex4}

\usepackage{graphicx}
\usepackage{dcolumn}
\usepackage{bm}
\usepackage{amsmath}

\newcommand{\beq}{\begin{eqnarray}}
\newcommand{\eeq}{\end{eqnarray}}

\newcommand{\pardis}{\langle {\cal M} \rangle}

\newcommand{\ie}{{i.e. }}

\newcommand{\real}{{\sf I}\kern-.12em{\sf R}}
\newcommand{\comp}{{\sf I}\kern-.50em{\sf C}}
\newcommand{\unity}{{\sf I}\kern-.54em{\sf 1}}

\newcommand{\tmtextrm}[1]{{\rmfamily{#1}}}

\def\spose#1{\hbox to 0pt{#1\hss}}
\def\ltapprox{\mathrel{\spose{\lower 3pt\hbox{$\mathchar"218$}}
 \raise 2.0pt\hbox{$\mathchar"13C$}}}

\begin{document}

\preprint{GEF-TH 12/07}

\title{Confining properties of QCD at finite
  temperature and density}
\author{Simone Conradi, Alessio D'Alessandro and Massimo D'Elia}
\affiliation{Dipartimento di Fisica dell'Universit\`a di Genova and INFN, Sezione di Genova, Via Dodecaneso 33, I-16146 Genova, Italy}

\date{\today}

\begin{abstract}
A disorder parameter detecting dual 
superconductivty of the vacuum is used as a probe to characterize
the confining properties of the phase diagram of two color
QCD at finite temperature and density. 
We obtain evidence for the disappearing of dual superconductivity
(deconfinement) induced by a finite density of baryonic matter, 
as well as for a
coincidence of this phenomenon with the restoration of chiral symmetry
both at zero and finite density.
The saturation transition induced by Pauli blocking 
is studied as well, and a general warning is given about the 
possible effects that this unphysical transition could have on
the study of the QCD phase diagram at strong values of the gauge
coupling.
\end{abstract}

\maketitle

\section{Introduction}

Color confinement emerges as an absolute property of strongly
interacting matter from experimental facts, but is not yet fully 
understood starting from the QCD first principles. Lattice QCD
simulations, however, have given some evidence about confinement
and have even predicted the presence of a finite 
temperature transition to a deconfined state of matter, which 
is currently under investigation in Heavy Ion Collision Experiments.
Deconfinement, has seen by  lattice simulations, appears to take place
at the same temperature where other important physical phenomena
happen, like for instance the restoration of chiral symmetry.

In the present study we address the question regarding the fate
of confining properties in presence of a finite density of baryonic
matter: that is important in order to understand the structure of
the QCD phase diagram both in the region of low densities and 
high temperatures, which is relevant for heavy ion experiments, 
and in the region of low temperatures and high densities, where
the study of confining properties could help 
characterizing the nature of matter in compact astrophysical objects.
In particular we are interested in understanding how
deconfinement induced by a critical density of baryonic matter 
compares to what happens at zero density,
and how it is related to other
possible transitions in the QCD phase diagram. 
Various possibilities are open in 
principle, like for instance a non-coincidence of the deconfinement
transition with the restoration of chiral symmetry.

The topic has been already studied in previous literature. 
In particular results about deconfinement at high
densities and low temperatures have been obtained in
Ref.~\cite{Hands06},
while the relation between deconfinement and the chiral transition
has been investigated in Refs.~\cite{immu_dl,topsu2}. 
However previous studies have been based on the study of the 
expectation value of the Polyakov loop as a probe for the 
deconfinement transition: while that can indeed be used as on order 
parameter for confinement in the pure gauge theory, being related to 
the spontaneous breaking of the center symmetry, the same is not true 
in the theory with dynamical fermions, where center symmetry
is explicitly broken. For that reason in the present study we
look for different order parameters which are constructed in 
the framework of specific mechanisms of color confinement and which
may be valid also in the full QCD theory.

One appealing mechanism, among others, 
is that based on dual superconductivity of 
the QCD vacuum, which relates confinement to the breaking of an abelian dual
symmetry induced by the condensation of magnetic 
monopoles~\cite{thooft75,mandelstam,parisi}.
The possibility to define disorder parameters in this scenario
has been studied since a long time, one parameter has been developed
by the Pisa group~\cite{del,DiGiacomo:1997sm}
and consists in the expectation value of an operator which creates a magnetic
monopole, $\pardis$\footnote{We change the usual notation for the disorder
  parameter, $\langle \mu \rangle$, 
in order to avoid confusion with the notation
  for the chemical potential.  
}: 
that has been shown to be a good parameter for 
color confinement both in pure Yang-Mills theories~\cite{PaperI,PaperIII}
and in full QCD~\cite{full1,full2}; similar parameters have been
developed elsewhere~\cite{moscow,bari,marchetti}.

Our plan is to use $\pardis$ as a parameter for  characterizing
the confining properties of the various phases in the QCD
phase diagram. Since numerical studies of QCD at finite density are notoriously
difficult because of the sign problem, which makes usual importance
sampling simulations unfeasible, in the present study we restrict
ourselves to the theory with two colors, where that problem is absent.
In principle no significant differences are expected for the confining
properties of the theory  when going from $N_c = 2$ to $N_c = 3$,
where $N_c$ is the number of colors: for that reason we believe that
our study could be relevant also for real QCD.

In Section~\ref{themodel} we recall the general properties of 
lattice QCD at finite baryon density as well as the specific features
of the two color model.
In Section~\ref{definition} we review 
the definition of the disorder parameter $\pardis$ and develop our
strategy to study its properties at finite density.
In Section~\ref{results} we present our numerical results: in
particular in Section~\ref{zeromu} we present a study of the 
disorder parameter 
at zero density; results at non-zero baryon density are given
in Section~\ref{nonzeromu}; a determination of the 
chiral transition line and its comparison with the deconfinement transition
line are presented in Section~\ref{critiline}. 
In Section~\ref{saturation} the disorder parameter $\pardis$
and other observables are 
used to study the nature of the unphysical transition 
to the saturation regime taking place at high values of the chemical
potential and the possible influence of saturation on the physical
transition is discussed. In Section~\ref{lowtemp} we discuss the
relevance of our results for the low temperature region of the 
phase diagram. Finally, in Section~\ref{conclusions}, we draw our conclusions.
A partial account of our results has been presented in 
Ref.~\cite{lat06}.

\section{QCD at finite density and the two-color model}
\label{themodel}

We will consider a discretized lattice action for two-color QCD
at finite chemical potential defined as follows:
\beq
S = S_G + \sum_{i,j} \bar\psi_i M [U]_{i,j} \psi_i
\label{action}
\eeq
where $S_G$ is the pure gauge Wilson action,
\beq
S_G = \beta \sum_{\square} \left(1 - {1 \over 2} {\rm Tr}\, \square
\right)\, , 
\eeq
the sum being over all plaquettes, while the fermion matrix is
defined, in the case of standard staggered fermions, as 
\begin{eqnarray}
M_{i,j} &=& a m
\delta_{i,j} + {1 \over 2} 
\sum_{\nu=1}^{3}\eta_{i,\nu}\left(U_{i,\nu}\delta_{i,j-\hat\nu}-
U^{\dag}_{i-\hat\nu,\nu}\delta_{i,j+\hat\nu}\right) \nonumber \\
&+& \eta_{i,4}
\left(e^{ a \mu}U_{i,4}\delta_{i,j-\hat4}-
e^{- a \mu}U^{\dag}_{i-\hat4,0}\delta_{i,j+\hat4}\right) \, .
\label{fmatrix}
\end{eqnarray}
Here $i$ and $j$ refer to lattice sites, $\hat\nu$ is a unit vector on
the lattice, $\eta_{i,\nu}$ are staggered phases 
and $U$ are gauge link variables;  $a \mu$ and $a m$ are respectively 
the chemical potential and the quark mass in lattice units.
The grand-canonical partition function can be written, after
integrating out fermions, as:
\begin{eqnarray}
Z=\int \mathcal{D}U
e^{-S_{G}[U]} \det M[U] \, .
\label{grandpart2}
\end{eqnarray}
In ordinary QCD the fermion determinant is complex for generic
values of the chemical potential, thus hindering the use 
of numerical Monte-Carlo simulations. Various possibilities
have been explored to circumvent the problem, like for instance
reweighting techniques~\cite{fodor,density},
the use of an imaginary chemical potential either 
for analytic continuation~\cite{muim,immu_dl,azcoiti,chen,giudice,cea,sqgp} 
or for reconstructing the canonical
partition function~\cite{cano}, Taylor expansion
techniques~\cite{taylor1,taylor2} and
non-relativistic expansions~\cite{hmass1,hmass2,hmass3}.

The problem is absent in QCD with two colors, since
the gauge group is real:
indeed the fermion determinant, being expressible like any other 
gauge invariant observable in terms of traces over closed loops,
is real as well, and numerical simulations are feasible.
For this reason two-color QCD has been widely studied
in the past as a laboratory for real QCD at finite 
density~\cite{Hands99,aloisio,Kogut01,scorzato,Kogut02,muroya,topsu2,Hands06,chandra}.
Despite some peculiar features of the model, like the fact
that baryons and mesons are degenerate, one still expects
to learn relevant information about specific questions, like
for instance the fate of topology~\cite{topsu2} or confinement at finite
density.

\section{The disorder parameter $\pardis$}
\label{definition}

The magnetically charged operator
{${\cal M} (\vec x,t)$}, whose expectation value detects dual
superconductivity, is defined in the continuum as the
operator which creates a magnetic monopole in $\vec x,t$ by shifting 
the quantum field by the classical vector potential of a monopole,
$\vec{b}_\perp$,
and can be written (see Ref.~\cite{DiGiacomo:1997sm} for details) as
\begin{equation}
{\cal M} ( \vec{x}, t) = \exp \left[\frac{i}{e} \int \text{\tmtextrm{d}}^3 y
  \hspace{0.25em} \vec{E}_\perp ( \vec{y}, t) \vec{b}_\perp ( \vec{y} - \vec{x}) \right]
  \, ,
\end{equation}
with the electric field $\vec{E}_\perp ( \vec{y}, t)$ being the
momentum conjugate to the quantum vector potential.
Its expectation value, when discretized on the lattice, can be
expressed as the ratio of two different partition functions,
\begin{eqnarray}
\label{defmu}
\pardis = \tilde{Z}/{Z} \, ,
\eeq
where $Z$ is the usual QCD partition function, while
$\tilde{Z}$ is obtained from $Z$ by a change in the pure
gauge action $S_G \to \tilde S_G$, consisting in the addition of the
monopole field to the temporal plaquettes  at a given timeslice 
where the monopole is created.

Being expressed as the ratio of two different partition functions, 
the numerical study of $\pardis$ is a highly non-trivial task,
since $\cal M$ gets
significant contributions only on those
configurations having very small statistical weight.
While numerical methods have been recently developed which 
permit a direct determination of $\pardis$~\cite{muu1}, we shall 
not use them in
the present study since they involve the combination of several
different Monte Carlo simulations, a task which in presence of dynamical
fermions could be unpractical. We will instead study, as usual, 
susceptibilities
of the disorder parameter, from which the behaviour of $\pardis$ at the 
phase transition can be inferred.

For instance, being interested in $\pardis$ as a function of $\beta$, 
as for the 
$\mu = 0$ phase transition, one usually 
measures~\cite{del,DiGiacomo:1997sm,PaperI}
\begin{eqnarray}
\rho = \frac{\partial}{\partial \beta} \ln \pardis = 
\frac{\partial}{\partial \beta} \ln \tilde Z -
\frac{\partial}{\partial \beta} \ln Z = \langle S \rangle_S -  \langle \tilde{S} \rangle_{\tilde{S}} \; 
\label{rhoferm}
\end{eqnarray}
where the subscript indicates 
the pure gauge action used for Monte Carlo sampling. 
The disorder parameter can be reconstructed from the susceptibility
$\rho$, exploiting the fact that one has exactly $\pardis = 1$ at $\beta = 0$  
\begin{eqnarray} 
\label{mufromrho}
\pardis (\beta) = \exp\left(\int_0^{\beta} \rho(\beta^{\prime}) {\rm d}
\beta^{\prime}\right) \; .
\end{eqnarray}
In particular $\rho \simeq 0$ in the confined phase means $\pardis
\neq 0$,
a  sharp negative peak of $\rho$ at the phase transition implies 
a sudden drop of $\pardis$ and $\rho$ diverging
in the thermodynamical limit in the deconfined phase means that 
$\pardis$ is exactly zero beyond the phase transition.

Studying $\pardis$ as a function of $\beta$ is what is usually done
if interested in the fate of dual superconductivity as the temperature
is increased.
Indeed, 
in the Euclidean path integral formulation of QCD, the physical
temperature is related to the inverse temporal extension, 
$T = 1/(L_t a)$,
where $a$ is the lattice spacing which for an asymptotically
free field theory is a decreasing function 
of the inverse gauge coupling $\beta$.
For that reason the inverse coupling $\beta$ is usually 
adopted in place of $T$ when studying the
QCD phase diagram, the latter being an increasing function
of the former.

At finite temperature and density we are interested in studying the 
behaviour of $\pardis$ in the two parameter
space $(\beta,\hat\mu)$, where $\hat\mu \equiv a \mu$ 
is the chemical potential in lattice units. For that
reason we introduce the new susceptibility

\begin{eqnarray}
\rho_D \hspace{-1pt}\equiv \hspace{-1pt}\frac{\partial}{\partial \hat\mu} \ln \pardis \hspace{-1pt}= \hspace{-1pt}
\frac{\partial \ln \tilde Z}{\partial \hat\mu} \hspace{-1pt} - \hspace{-1pt}
\frac{\partial \ln Z}{\partial \hat\mu}  \hspace{-1pt} = \hspace{-1pt}
\langle N_q \rangle_{\tilde S} -  \langle N_q \rangle_{{S}} \; 
\label{rhod}
\end{eqnarray}
where $N_q$ is the quark number operator, \ie according to the
definition of $Z$ given in Eq.~(\ref{grandpart2}):  
\beq
 \langle N_q \rangle = \left\langle \, {\rm Tr} \left( \frac{\partial
   M} {\partial
  \hat\mu} \cdot  M^{-1} \right) \right\rangle;
\eeq 
(an additional factor 2 is actually needed for the case studied
in the present paper, which deals with 8 staggered flavors, see
Eq.~(\ref{grandpart3})).
The dependence of $\pardis$ on the chemical potential $\mu$ can then be 
reconstructed as follows:
\beq
\pardis (\beta,\hat\mu) = \pardis (\beta, 0) \exp\left(\int_0^{\hat\mu}
\rho_D(\hat\mu^{\prime}) {\rm d} \hat\mu^{\prime}\right) \; \nonumber
\label{recond}
\eeq
so that, if the starting point at $\hat\mu = 0$ is in the confined phase
($\pardis (\beta, 0) \neq 0$), 
the behaviour expected for $\rho_D (\hat\mu)$ in correspondence of a
possible finite density deconfinement transition will be the same shown
by $\rho$ across the finite temperature transition.

Assuming the presence of a (pseudo)critical line in the $T - \mu$
plane where the disorder parameter drops to zero and dual superconductivity
disappears, the two susceptibilities $\rho$ and $\rho_D$ can be used
not only to locate the position of the line, but also to compute its 
slope, thus providing a more comprehensive information about the 
QCD phase diagram. Indeed, it is quite natural to assume
that the gradient of the disorder parameter, 
\beq
\vec\nabla\pardis = 
\left( {\partial \pardis \over \partial \beta},\, 
{\partial \pardis \over \partial \hat\mu}
 \right) = (\rho,\, \rho_D) \, \pardis ,
\eeq
be orthogonal, in the $\beta - \hat\mu$ plane, to the critical line, 
whose slope is then equal to $- \rho_D/\rho$. In the following we shall
directly check this property on our numerical data and also make use
of it to obtain testable predictions.

\section{Numerical results}
\label{results}

In order to perform numerical simulations we have adopted the usual
Hybrid Monte Carlo algorithm. The partition function in 
Eq.~(\ref{grandpart2}) can be rewritten, introducing pseudo-fermionic
fields $\Phi$, as 
\begin{eqnarray}
Z&=&\int \mathcal{D}U\mathcal{D}\Phi
e^{-S_{g}[U]-\Phi^{*}(M^{\dag}M)^{-1}\Phi} \nonumber \\ 
&=& 
\int \mathcal{D}U
e^{-S_{g}[U]} \left( \det M[U] \right)^2 \, .
\label{grandpart3}
\end{eqnarray}
In presence of a real chemical potential the usual
even-odd factorization trick for reducing the number of flavors
cannot be performed, so that Eq.~(\ref{grandpart3}) actually
describes a theory with 8 (degenerate in the continuum limit)
flavours. The standard exact $\phi$ algorithm described in 
Ref.~\cite{Gottlieb:1987mq} has been used.

We have performed simulations on lattices $L_s^3 \times L_t$ with
$L_t = 6$ and different values of the spatial size ranging from
$L_s = 8$ to $L_s = 16$. The bare quark
mass has been fixed to $a m = 0.07$.

Simulations on the smallest lattice ($L_s = 8$) have been performed
on a PC farm, making use of a numerical code obtained by adapting
the publicly available MILC code for two colors and for 
the inclusion of a finite chemical potential.
Simulations on larger lattices have been performed instead 
on the INFN apeNEXT facility in Rome.

The observables we look at are, apart from the susceptibilities of the 
disorder operator introduced in Section~\ref{definition}, 
the average Polyakov loop, the average plaquette and the chiral 
condensate: 
\beq
\langle L \rangle 
\equiv \frac{1}{L_s^{3}} \sum_{n} \frac{1}{N_c} \langle {\rm Tr}\
       {\bf L} ({n}) \rangle \, ,
\eeq

\beq
\langle P \rangle 
\equiv \frac{1}{6 L_t L_s^{3}} 
\sum_{n, \mu < \nu} \frac{1}{N_c} \langle {\rm Tr}\ \Pi_{\mu\nu} (n)
\rangle \, , 
\eeq

\beq
\langle \bar \psi \psi \rangle 
\equiv \frac{1}{L_t L_s^{3}} \langle {\rm Tr}\ M^{-1} \rangle \, ,
\eeq

\noindent
as well as their susceptibilities

\beq
\chi_c \equiv L_s^3 L_t\ \langle 
(\bar \psi \psi - \langle \bar \psi \psi\rangle)^2 
\rangle \, , 
\eeq

\beq
\chi_L \equiv L_s^3\ \langle 
(L - \langle L \rangle)^2 
\rangle \, , 
\eeq

\beq
\chi_P \equiv L_s^3 L_t\ \langle 
(P - \langle P \rangle)^2 
\rangle \, .
\eeq
Notice that in the case of the chiral susceptibility we have
explicitly considered only the disconnected contribution.

\subsection{The deconfining transition at zero chemical potential}
\label{zeromu}

It is a well known fact that in ordinary full QCD
at zero baryon density, 
chiral symmetry restoration takes place at the same 
critical temperature as deconfinement, with the latter identified with
the disappearance of dual superconductivity~\cite{full1,full2}.
We will check again this fact for the theory with two colors,
since this will be an important reference information for our
following analysis at finite density.

\begin{figure}[h!]
\includegraphics*[width=1.0\columnwidth]{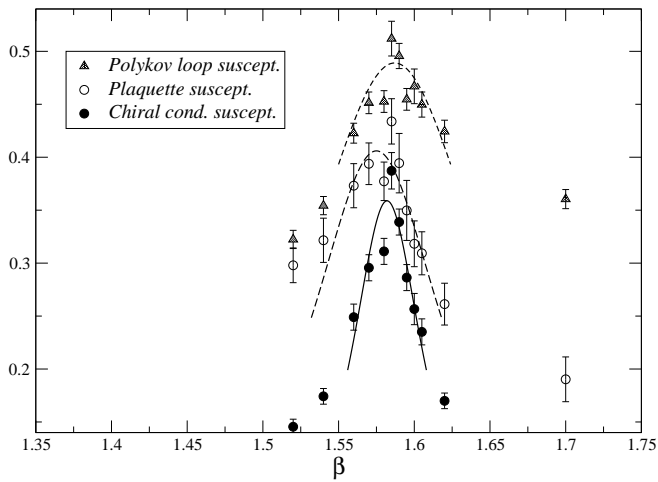}
\vspace{-0.cm}
\caption{Plaquette, Polyakov and chiral susceptibilities on the
$16^3 \times 6$ lattice at $\hat\mu = 0$. 
The chiral and the plaquette susceptibility 
have been respectively divided by a factor 10 and multiplied by a
factor 4 in order to fit in the figure. Curves corresponding to best
fits of the peak positions are superposed to the numerical data.}
\label{fig1} 
\vspace{-0.cm}
\end{figure}

We show in Fig.~\ref{fig1} the peaks of the three susceptibilities
$\chi_c$, $\chi_L$ and $\chi_P$ defined above, obtained on
a $16^3 \times 6$ lattice, together with curves
corresponding to best fits to the location of their peaks. 
Our estimate for the location of the transition, obtained through
a fit to the chiral susceptibility, is $\beta_c = 1.582(2)$, 
to be compared to those obtained by fitting the Polyakov
loop susceptibility ($\beta_L = 1.587(4)$) and the plaquette
susceptibility ($\beta_P = 1.575(5)$). A clear drop of the chiral
condensate and a rise of the Polyakov loop are also observed 
at $\beta_c$, as shown in Fig.~\ref{polychi}.
The dependence of $\beta_c$ on the spatial size is not significant,
as can be appreciated from Table~\ref{couplings}, 
where we report a summary of the 
pseudo-critical couplings (and chemical potentials) obtained
from our simulations.

\begin{figure}[t!]
\hspace{0.4cm}\includegraphics*[width=0.95\columnwidth]{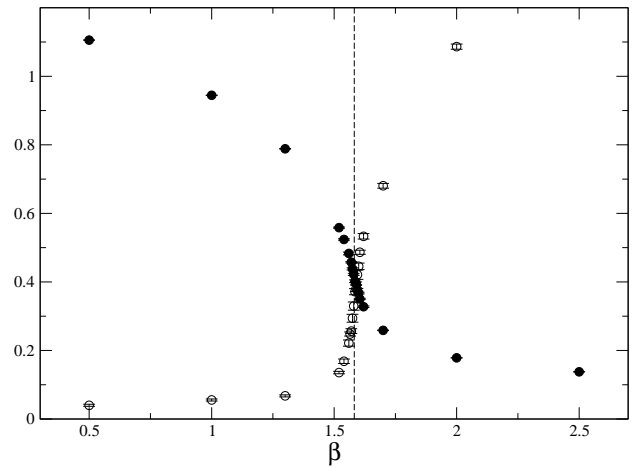}
\vspace{-0.cm}
\caption{Chiral condensate and average Polykov loop
as a function of $\beta$ measured on 
a $16^3 \times 6$ lattice at $\hat\mu = 0$. The vertical line indicates the
pseudo-critical coupling $\beta_c$ as determined from a best fit
to the peak of the chiral susceptibility. The Polyakov loop
has been multiplied by a factor 15 the better fit in the figure.}
\label{polychi} 
\vspace{-0.cm}
\end{figure}

\begin{figure}[t!]
\includegraphics*[width=1.0\columnwidth]{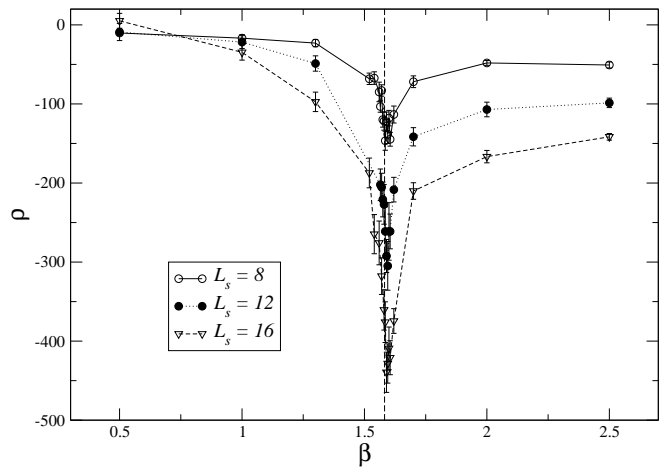}
\vspace{-0.cm}
\caption{$\rho$ parameter as a function of $\beta$
on various lattice sizes. The vertical line corresponds
again to the location of the chiral transition.}
\label{rhobeta} 
\vspace{-0.cm}
\end{figure}

Let us now consider the fate of dual superconductivity.
In Fig.~\ref{rhobeta} we show the behaviour of the susceptibility
$\rho$ as a function of $\beta$ for three different lattice sizes.
A clear peak can be appreciated, which deepens when increasing
the lattice size and whose location is clearly coincident with that
of the chiral transition.
Moreover it is also apparent from the figure that $\rho$ is practically
independent of the lattice size in the low coupling region, confirming
that $\pardis \neq 0$ in the thermodynamical limit in that phase, 
while $\rho$ strongly depends on $L_s$, and in particular is linear
with it, as shown in Fig.~\ref{fig:rhohb}, in the weak coupling region, 
showing that $\pardis$ is exactly equal to zero in the thermodynamical
limit beyond the transition (magnetic charge 
superselection~\cite{supersel}). Therefore $\beta_c$ seems to separate 
two phases characterized by a different realization of the 
$U(1)$ magnetic symmetry. 

\begin{figure}[t!]
\includegraphics*[width=1.0\columnwidth]{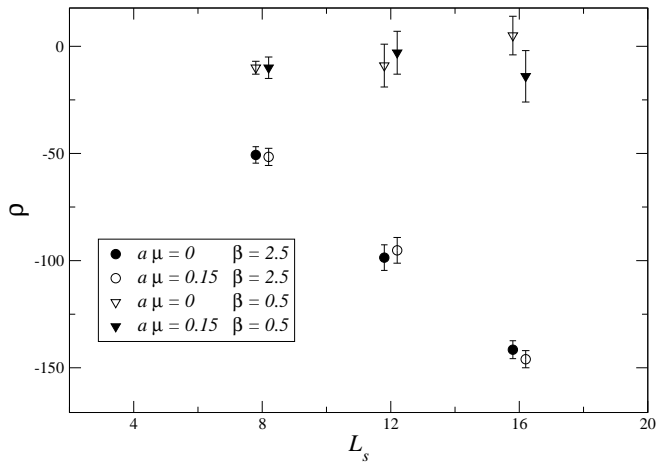}
\vspace{-0.cm}
\caption{Behaviour of $\rho$ in the strong coupling
($\beta = 0.5$) and in the weak coupling ($\beta = 2.5$) region
as a function of $L_s$ for $\hat\mu = 0$ and $\hat\mu = 0.15$.
$\rho$ stays constant and close to zero in the thermodynamical 
limit at strong coupling, while it diverges linearly with $L_s$ 
at weak coupling.}
\label{fig:rhohb} 
\vspace{-0.cm}
\end{figure}

To better appreciate the coincidence of the chiral transition
with the disappearance of dual superconductivity, we have tried
a finite size scaling (f.s.s.) analysis of the critical behaviour of $\pardis$
around the transition temperature. 
We can assume for $\pardis$ the following f.s.s. ansatz: 
\beq
\pardis = L_s^{- \frac{\delta}{\nu}} \phi \left( (\beta_c - \beta)
  L_s^{1/\nu} \right)
\eeq
from which it can be easily derived
\beq
\rho = L_s^{1/\nu} \tilde\phi((\beta_c - \beta) L_s^{1/\nu}) \; .
\eeq
We have checked this ansatz on our data,
obtaining  the best possible agreement for $\nu \simeq 0.63$
and $\beta_c \simeq 1.584$: a reasonable scaling is obtained, with
deviations observed on the smaller lattice. In particular we estimate
$\beta_c = 1.584(2)$ in good agreement with the 
location of the chiral transition given above.
The fitted critical index $\nu$ seems to indicate an 
Ising 3D critical behaviour, to be compared to that 
taking place in the quenched limit
(3D Ising) and the renormalization group 
prediction for the critical behaviour in 
the chiral limit (first order~\cite{wirstam}). 
However a similar finite size scaling is not observed 
for the other susceptibilities and we believe
that a definite answer about the universality class
of the transition cannot be given in the present context,
also due to the relatively small spatial volume used
(our largest aspect ratio is slightly less than 3).
A more careful investigation should be performed and we consider
the present analysis, as well as that presented later 
for the finite density case, as only 
aimed at a quantitative estimate of the critical coupling
where superconductivity disappears.

\begin{figure}[t!]
\includegraphics*[width=1.0\columnwidth]{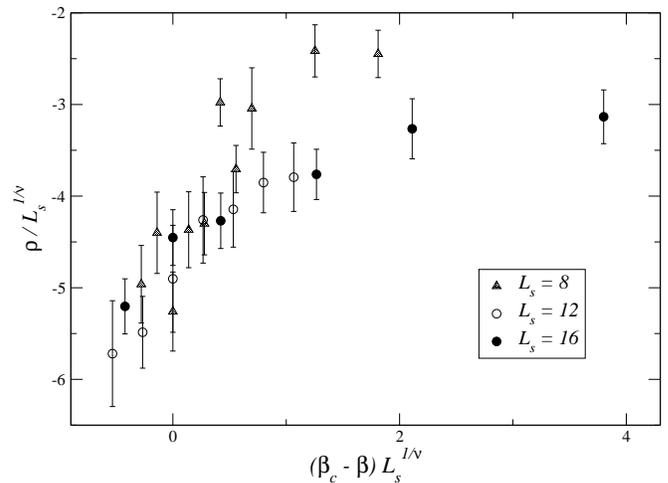}
\vspace{-0.cm}
\caption{Finite size scaling analysis of $\rho$ around the
  transition. The best possible scaling is obtained for
$\nu \simeq 0.63$ and $\beta_c \simeq 1.584$.}
\label{fig:fss} 
\vspace{-0.cm}
\end{figure}

\subsection{The deconfining transition at non zero chemical potential}
\label{nonzeromu}

The two susceptibilities $\rho$ and $\rho_D$ permit to study 
$\pardis$ either as a function of temperature at a fixed value of 
the chemical potential $\hat\mu$, or as a function of $\hat\mu$
at fixed temperature. Both strategies can be used to investigate
the fate of dual superconductivity in presence of a finite density 
of baryonic matter: the first could be more effective 
at small chemical potentials, where the possible transition
line starting at $\hat\mu = 0$ 
should be almost parallel to the $\hat\mu$ axis, the second
could be more convenient at larger chemical potentials.
Actually a proper combination of $\rho$ and $\rho_D$ could be used to 
study the behaviour of $\pardis$ along any given path
in the $\beta - \hat\mu$ plane so that one could even choose an optimal 
combination corresponding to a relevant direction around a critical point:
however we shall limit ourselves in the present context to the simpler cases
of either fixed temperature or fixed chemical potential.
The study at fixed temperature has a particular interest, since
it may show how the disappearance of confinement (dual
superconductivity) can be induced by simply increasing the 
density of baryonic matter.

We shall first consider the case of a fixed chemical potential,
$\hat\mu = 0.15$. In Fig.~\ref{fig:chisuscbeta} we show the chiral
susceptibility obtained on a $16^3 \times 6$ lattice and compared
to the same quantity computed at $\hat\mu = 0$. A clear shift 
of the pseudocritical coupling can be appreciated, in particular
we obtain $\beta_c(\hat \mu = 0.15) = 1.568(2)$, showing that the
(pseudo)critical temperature lowers as the chemical potential is
increased.
Data for the susceptibility $\rho$ on the same lattice are shown in 
Fig.~\ref{fig:rho_beta} and compared to those obtained 
at zero density: the peak of $\rho$ shifts consistently by an amount
comparable to that of the chiral susceptibility. 
Notice that in both cases the actual position of the $\rho$ 
peak is at a $\beta$ slightly larger than $\beta_c$. That is expected
since $\rho$ is a logarithmic derivative:
assuming that 
$\pardis' \equiv \partial \pardis / \partial \beta$ has a minimum
at $\beta_c$, it follows that 
$\partial \rho / \partial \beta = \pardis''/\pardis -
(\pardis'/\pardis)^2$ is still negative at the same point.

Data reported in Fig.~\ref{fig:rhohb} show that, 
also in the case $\hat\mu = 0.15$, $\rho$ is independent of the lattice size 
and practically equal to zero in the strong coupling region,
while it diverges linearly with $L_s$ in the weak coupling region.
Therefore we can conclude that, also in presence of a finite density
of baryonic matter, dual superconductivity disappears as the
temperature is increased at the same point where chiral symmetry
is restored.

\begin{figure}[t!]
\includegraphics*[width=1.0\columnwidth]{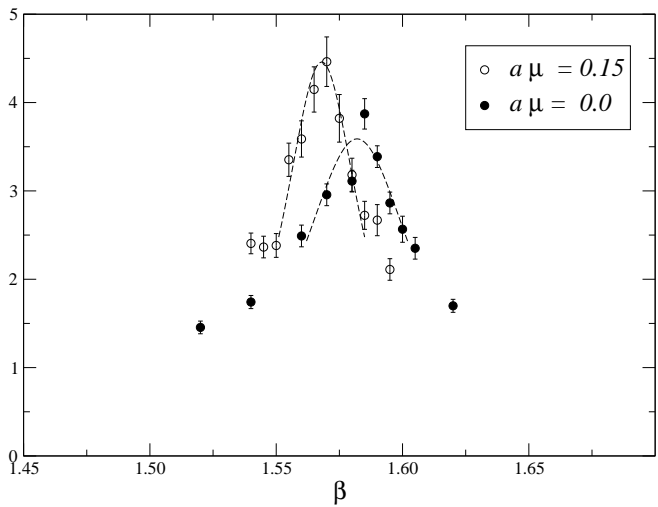}
\vspace{-0.cm}
\caption{Chiral susceptibility on a $16^3 \times 6$ 
lattice as a function of $\beta$ for various
values of $\hat\mu$. Dotted curves correspond to best fit to the peak
values.}
\label{fig:chisuscbeta} 
\vspace{-0.cm}
\end{figure}

\begin{figure}[b!]
\includegraphics*[width=1.0\columnwidth]{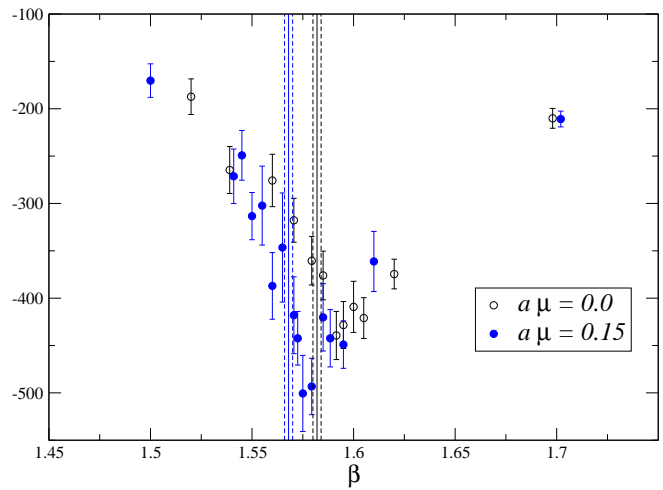}
\vspace{-0.cm}
\caption{$\rho$ on a $16^3 \times 6$ lattice as a function of $\beta$ 
for various values of $\hat\mu$. Vertical bands correspond to the 
pseudo-critical $\beta$ fitted according to 
the chiral susceptibility.}
\label{fig:rho_beta} 
\vspace{-0.cm}
\end{figure}

Next we turn to the behaviour of $\pardis$ as a function 
of $\hat\mu$ at fixed temperature ($\beta$), determined by means
of the susceptibility $\rho_D$. We have considered only
values of $\beta$ below the (pseudo)critical coupling
$\beta_c$ computed at $\hat\mu = 0$, in particular $\beta = 1.50$
and $\beta = 1.55$:
 in this case we know that 
$\pardis \neq 0$ at $\hat\mu = 0$, so that $\rho_D$ may signal a
possible disappearance of dual superconductivity induced by 
finite baryon density.
Notice that the lowest value of $\beta$,
on the basis of a rough two-loop estimate of the 
$\beta$-function, corresponds to a physical temperature
$T/T_c \sim a(\beta = 1.582)/a(\beta = 1.5) \sim 0.4$,
where $T_c$ is the critical temperature at zero chemical
potential.

In Fig.~\ref{fig:chisuscmu} we show the chiral susceptibility
determined on a $16^3 \times 6$ lattice at $\beta = 1.55$ 
and $\beta = 1.50$. 
A best fit permits to locate the peak 
positions, hence the (pseudo)critical values of $\hat\mu$
corresponding to chiral restoration. We obtain
$\hat\mu_c (\beta = 1.50) = 0.340(10)$\footnote{
Notice that the (pseudo)critical chemical potential obtained 
at $\beta = 1.50$ is different from what obtained in 
Ref.~\cite{topsu2}: this difference can be understood in terms
of the different algorithm used, which in the case of 
Ref.~\cite{topsu2} was a non-exact molecular dynamics algorithm. 
}
and 
$\hat\mu_c (\beta = 1.55) = 0.215(10)$, as also reported in 
Table~\ref{couplings}.

\begin{figure}[t!]
\includegraphics*[width=1.0\columnwidth]{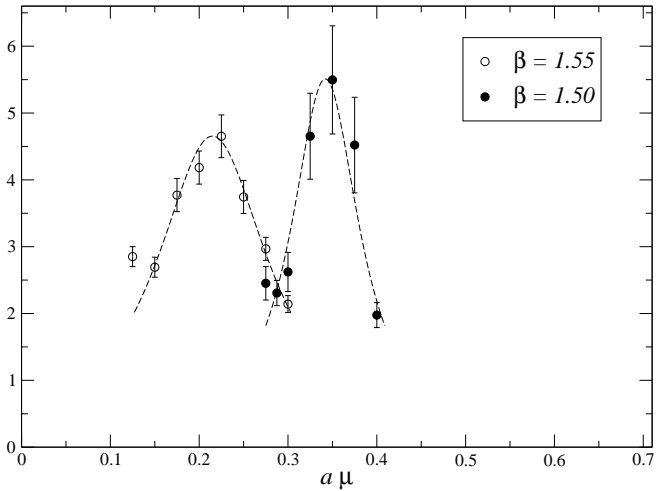}
\vspace{-0.cm}
\caption{Chiral susceptibility on a $16^3 \times 6$ lattice
as a function of $\hat\mu$ for various
values of $\beta$. Dotted curves correspond to best fit to the peak
values.}
\label{fig:chisuscmu} 
\vspace{-0.cm}
\end{figure}

\begin{figure}[t!]
\includegraphics*[width=1.0\columnwidth]{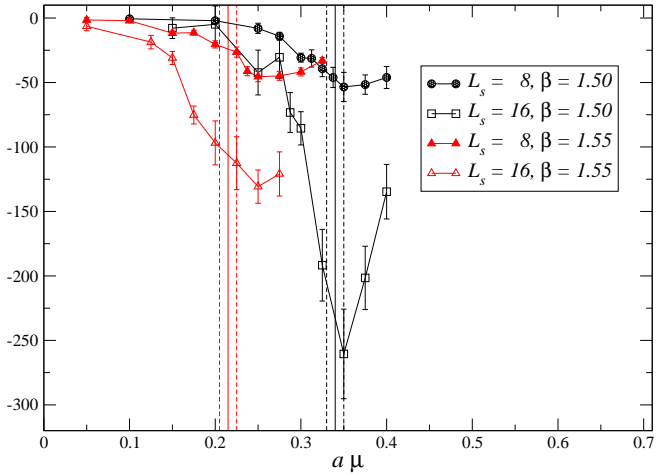}
\vspace{-0.cm}
\caption{$\rho_D$ as a function of $\hat\mu$ and for two lattice sizes
for various values of $\beta$. Vertical bands correspond to the 
pseudo-critical chemical potential fitted according to 
the chiral susceptibility.}
\label{fig:rhod_mu} 
\vspace{-0.cm}
\end{figure}

In Fig.~\ref{fig:rhod_mu} we show instead the results obtained for 
$\rho_D$ as a function of $\hat\mu$ at the same values of $\beta$ 
and on various lattice sizes. It clearly appears that while 
$\rho_D$ is independent of the lattice size and practically 
vanishing for small chemical potentials, it has a sharp negative 
peak in correspondence of the chiral transition which deepens
as the spatial size is increased. 
In order to be more quantitative about
the coincidence of chiral restoration and deconfinement, 
we have performed a f.s.s. analysis for the case
$\beta = 1.50$, where three different spatial sizes
were available ($L_s = 8,12,16$), according to the ansatz
\beq
\pardis = L_s^{- \frac{\delta}{\nu}} \phi \left( (\hat\mu_c - \hat\mu)
  L_s^{1/\nu} \right)
\eeq
hence
\beq
\rho_D = L_s^{1/\nu} \tilde\phi((\hat\mu_c - \hat\mu) L_s^{1/\nu}) \; .
\eeq
A reasonable scaling is obtained for 
$\nu \sim 0.55$ and $\hat\mu_c \sim 0.31$, in particular
we estimate $\beta_c = 0.315(15)$, marginally compatible
with the location of the chiral transition.

\begin{figure}[h!]
\includegraphics*[width=1.0\columnwidth]{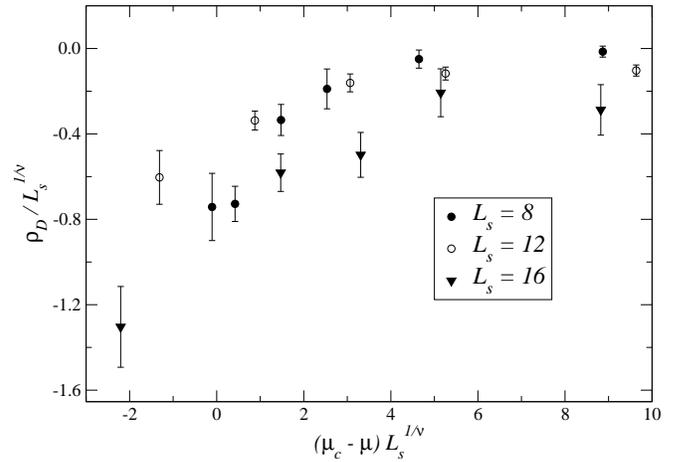}
\vspace{-0.cm}
\caption{Finite size scaling analysis for $\rho_D$. A critical index
$\nu \sim 0.55$ has been used,
the best value for the critical chemical potential being
$\beta_c \simeq 0.315(15)$.} 
\label{fig:fssmu} 
\vspace{-0.cm}
\end{figure}

We can therefore draw two important conclusions: dual
superconductivity (confinement) disappears in presence of a critical
density of baryonic matter; moreover the critical line in 
the $T - \mu$ plane corresponding
to deconfinement coincides, at least within our present uncertainties,
with the chiral transition line.
These results concern that part of the phase 
diagram including temperatures down to $T/T_c \sim 0.4$,
where $T_c$ is the critical temperature at zero chemical
potential: we shall discuss their relevance for the $T \sim 0$
region of the phase diagram later in this paper.

\subsection{The transition line}
\label{critiline}

Having obtained four different locations of the 
transition line, 
in particular $\beta_c ( \hat\mu = 0) = 1.582(2)$, 
$\beta_c ( \hat\mu = 0.15) = 1.568(2)$, 
$\hat\mu_c ( \beta = 1.55) = 0.215(10)$ and
$\hat\mu_c ( \beta = 1.50) = 0.340(10)$, as obtained
on our larger lattices (see Table~\ref{couplings}), we can perform
a fit of the dependence $\beta_c (\mu)$ in the whole 
$\beta - \hat\mu$ plane, which will then be used in the following.
We are also interested in testing what stated  
in Section~\ref{definition}, \ie that the ratio 
$ - \rho_D/\rho$ at the transition point can be used as an 
estimate of the slope of the critical line: we give an example
of a common plot of the two susceptibilites in Fig.~\ref{fig:rhorhod},
from which the ratio at $\beta_c,\hat\mu_c$ can be inferred.

\begin{figure}[h!]
\includegraphics*[width=1.0\columnwidth]{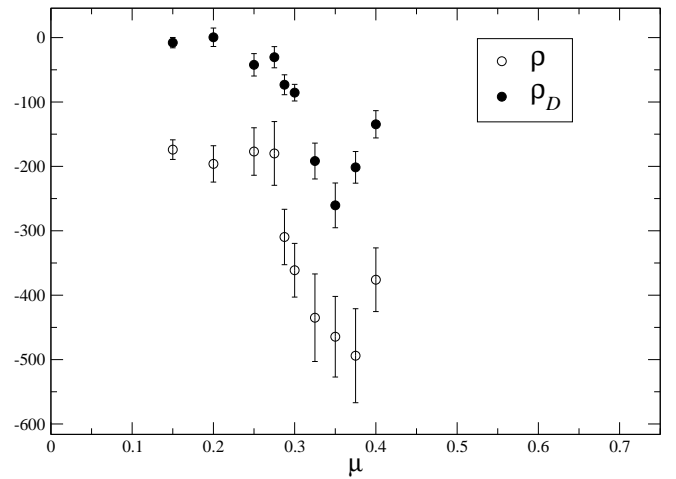}
\vspace{-0.cm}
\caption{
Comparison of $\rho$ and $\rho_D$ as a function of
$\hat\mu$ at $\beta = 1.50$ on a $16^3
\times 6$ lattice.} 
\label{fig:rhorhod} 
\vspace{-0.cm}
\end{figure}

We have tried a quadratic fit $\beta_c(\hat\mu) = A + B \hat\mu^2$,
obtaining $A = 1.5828(16)$, $B = -0.071(4)$ 
and $\chi^2/{\rm d.o.f.} = 0.26$.
The good value of $\chi^2/{\rm d.o.f.}$ shows that a quadratic
dependence well describes the critical line down to 
$T/T_c \sim 0.5$; indeed a fit with a quartic term gives 
a coefficient for $\hat\mu^4$ compatible with zero.
Our estimates for the location of the (pseudo)critical
points are reported in Fig.~\ref{diagr} together with the
fitted transition line. 

In correspondence of our direct locations of the transition
line we also show the estimates for the slope of the 
line obtained from the ratio $ - \rho_D/\rho$: in particular
we have drawn angles corresponding to one standard deviation
from the average values. A good agreement can be appreciated,
showing that $ - \rho_D/\rho$ can indeed be taken as a good
estimator of the slope of the line in 
the $\beta - \hat\mu$ plane.

\begin{figure}[t!]
\includegraphics*[width=1.05\columnwidth]{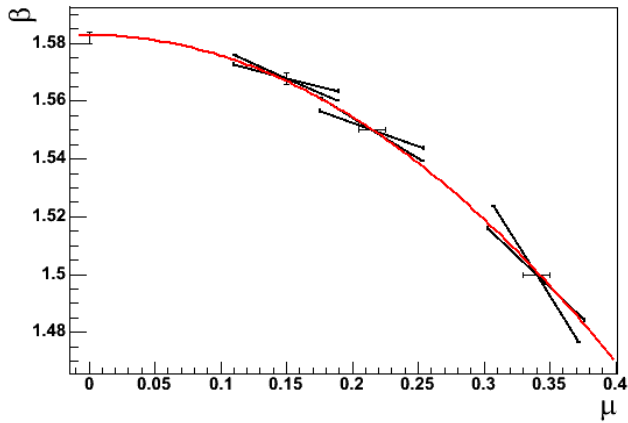}
\vspace{-0.cm}
\caption{Phase diagram in the $\beta - \mu$ plane. The chiral line
has been fitted to a quadratic dependence on $\mu$. The 
slope of the critical line, has inferred from the disorder 
parameter for dual superconductivity, has been reported
in the figure: a nice agreement (within one standard deviation)
can be appreciated.} 
\label{diagr} 
\vspace{-0.cm}
\end{figure}

\begin{figure}[b!]
\includegraphics*[width=1.0\columnwidth]{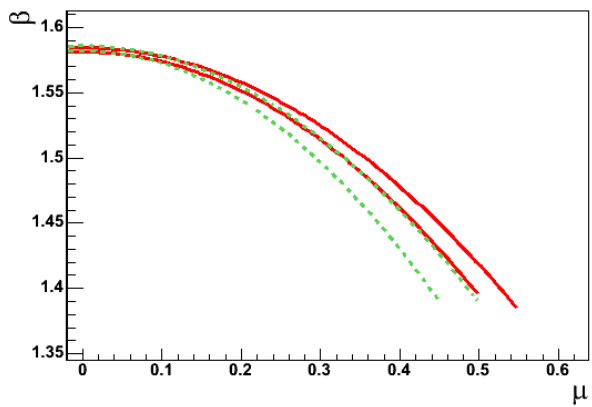}
\vspace{-0.cm}
\caption{Comparison of the chiral pseudocritical line (continuous)
and of that
corresponding to the disappearance of dual superconductivity (dotted), 
as
fitted from our data.}
\label{compareline} 
\vspace{-0.cm}
\end{figure}

Finally, in Fig.~\ref{compareline}, we report again
the chiral transition line fitted above and compared to a
quadratic fit in $\hat\mu$ for the critical line corresponding
to the disappearance of dual superconductivity (deconfinement).
The plot supports our previous statement, \ie that the chiral
transition coincides with deconfinement in the range
of $\beta$ values (temperatures) explored.

\subsection{A few remarks on saturation} 
\label{saturation}

It is a well known fact that, 
even in absence of the sign problem,
the study of lattice gauge theories in presence of a finite density
of fermions cannot be pushed to arbitrarily high densities,
 \ie to arbitrarily high values of the chemical
potential. Indeed the number of available energy levels
is limited by the presence of the UV cutoff, which places
an upper limit to the possible values of the Fermi energy. Stated 
otherwise, we cannot place, because of the Pauli exclusion 
principle,  more than one fermion with given quantum
numbers per lattice site. Apart from the upper limit 
that this places on the densities reachable on the lattice,
a much worse problem comes from the fact that, as saturation sets
in, the absence of available fermion levels quenches fermion 
dynamics, modifying the field theory at the ultraviolet scale.
As a matter of fact, the theory becomes equivalent to a pure
gauge theory in the large $\hat\mu$ limit.

Saturation is therefore an unphysical lattice artifact 
which may in principle invalidate numerical results,
one should therefore be extremely careful in locating 
its onset. Indeed, while saturation effects
are generically expected to appear for
$\hat\mu = a \mu$ of order 1, the exact value of $\hat\mu$
where they start to be important may depend on the dynamics
of the theory. In the following we will briefly explore
the transition to saturation in the two color model under 
consideration, arriving to some interesting conclusions
which may sound as a general warning.

We have explored saturation effects in some detail 
at $\beta = 1.55$. In Fig.~\ref{fig:saturobs} we show 
the behaviour of some observables as a function of $\hat\mu$ 
in a wide range going up to $\hat\mu = 1.6$.
For small values of the chemical potential above $\hat\mu_c$ the
fermion density rises roughly with a cubic dependence in $\hat\mu$,
as expected for a gas of free fermions,
but then saturates to a value which in the figure is normalized
to two fermions per site: the departure from the cubic behaviour
starts at $\hat\mu \sim 0.6 - 0.8$. Also the rise of the 
Polyakov loop suddenly stops at a similar value of 
$\hat\mu$, followed by a drop; in the same region the plaquette
suddenly drops towards its quenched value. Complete
saturation is reached for $\hat\mu \sim 1.4 - 1.6$.

Much is learned by looking at the behaviour of the susceptibilities
of the disorder parameter in the same range, which is shown in 
Fig.~\ref{fig:saturmu}: the negative peak of $\rho_D$ 
at $\hat\mu \sim 0.3$, corresponding to the physical deconfinement
transition, is followed by a positive unphysical peak 
at $\hat\mu \sim 0.7$. That means that the disorder parameter
$\pardis$, which at first
drops to zero thus signalling deconfinement, then rises
again as an effect of saturation: indeed the 
``saturation transition'' leads to a $SU(2)$ pure gauge theory,
which at $\beta = 1.55$ and $L_t = 6$ is deep in the confined
phase, implying $\pardis \neq 0$. To verify that we have explicitly 
reconstructed 
$\pardis(\hat\mu)/\pardis(\hat\mu = 0)$ (see Eq.~\ref{recond}) and reported
it in Fig.~\ref{fig:saturmu}: in the same figure we have reported
the location of the saturation transition as obtained by a fit
to the peak of the plaquette susceptibility.

We should be satisfied, since the saturation transition at 
$\hat\mu \sim 0.7$ is well separated from the physical transition
at $\hat\mu \sim 0.3$. However we notice that, defining a 
``saturation line'' in the
$\beta - \hat\mu$ plane corresponding to the onset of saturation
effects, we can predict, according to what stated in the previous
paragraph, its slope from the ratio $ - \rho_D/\rho$. We see from
Fig.~\ref{fig:saturmu}) that in correspondence
of the positive saturation peak for $\rho_D$, the other susceptibility
$\rho$ has a negative peak, hence we expect a positive slope
for the saturation line. That means that at lower values
of $\beta$ the onset of saturation
could take place at lower values of $\hat\mu$: that combined with 
the fact that the physical critical $\hat\mu_c$ instead increases as $\beta$
decreased, could lead to the unfortunate situation in which the
two transition, physical and unphysical, merge at lower values of $\beta$,
thus hindering, at least in the present case, the study of the strong
coupling (low temperature) region of the phase diagram.

\begin{figure}[h!]
\includegraphics*[width=1.0\columnwidth]{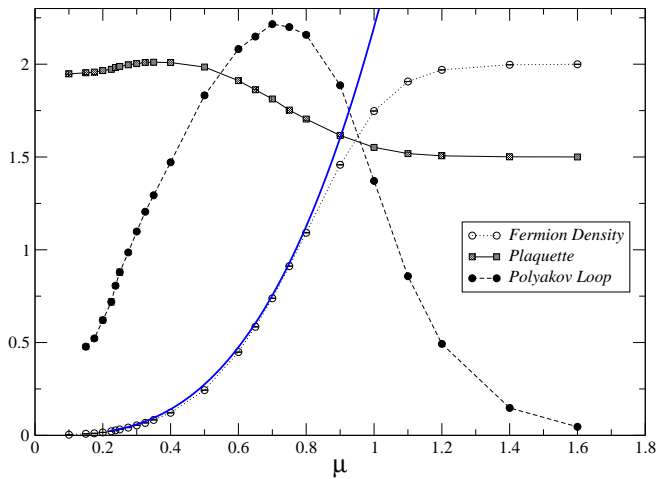}
\vspace{-0.cm}
\caption{
Various observables computed at 
$\beta = 1.55$ on the $8^3
\times 6$ lattice to show saturation. The Polyakov loop has been
multiplied by a factor 8, the average plaquette by a factor 4.} 
\label{fig:saturobs} 
\vspace{-0.cm}
\end{figure}

\begin{figure}[h!]
\includegraphics*[width=1.0\columnwidth]{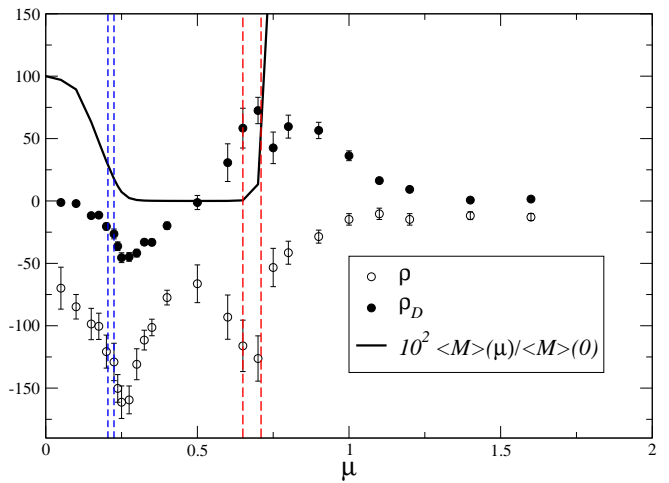}
\vspace{-0.cm}
\caption{
$\rho$ and $\rho_D$ at saturation. Notice the inversion of the 
peaks, indicating a transition line with positive slope.
Vertical dotted bands refer to the locations of the physical
and saturation transition.
The thick continuous line refers to the disorder parameter
$\pardis$ reconstructed by using the susceptibility $\rho_D$:
after an intermediate region where the magnetic symmetry
is restored, dual superconductivity sets in again in correspondence
of saturation.}
\label{fig:saturmu} 
\vspace{-0.cm}
\end{figure}

\begin{figure}[t!]
\vspace{0.5cm}
\includegraphics*[width=1.0\columnwidth]{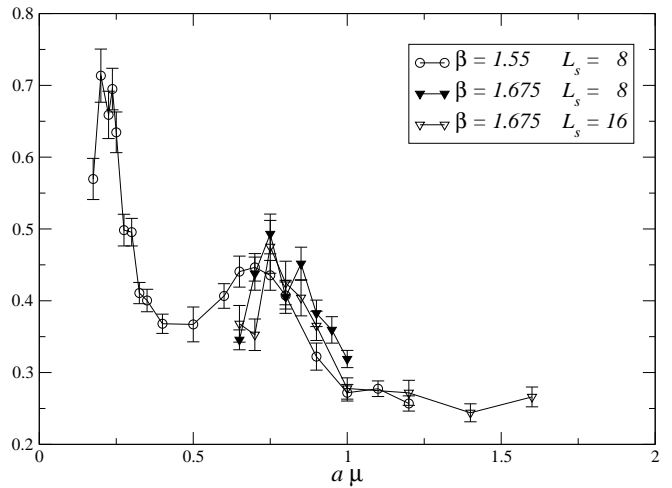}
\vspace{-0.cm}
\caption{Plaquette susceptibilities used to locate the saturation
transition.}
\label{plaqsat} 
\vspace{-0.cm}
\end{figure}

\begin{figure}[t!]
\vspace{0.5cm}
\includegraphics*[width=1.05\columnwidth]{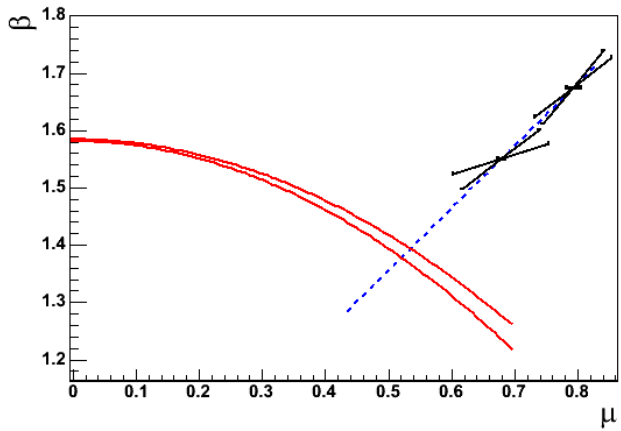}
\vspace{-0.cm}
\caption{Saturation transition line and its relation with the physical
transition line.}
\label{diagrsat} 
\vspace{-0.cm}
\end{figure}

In order to further explore this possibility we have decided
to make an estimate of the location of the saturation
transition, through a fit to the plaquette susceptibilities
(which are reported in Fig.~\ref{plaqsat}), performing
simulations also at a different value of the gauge coupling,
$\beta = 1.675$. Our estimate for the 
psudocritical saturation chemical potential $\mu_{Sc}$ are reported
in Table~\ref{couplings} and are
$\hat\mu_{Sc} (\beta = 1.55) = 0.68(3)$ and 
$\hat\mu_{Sc} (\beta = 1.675) = 0.79(3)$.
In Fig.~\ref{diagrsat} we report our estimate for the location
of the saturation line together with a rough linear
extrapolation suggesting that the saturation line could 
meet the physical line, whose estimate given in previous paragraph
is reported in the figure as well, for $\beta \sim 1.4$. 
Notice that the linear extrapolation adopted is 
supported by the slope of the line 
obtained through the ratio 
$- \rho_D/\rho$, whose estimates 
are reported in the figure as well.

We therefore give a general warning about the possible effects
of saturation on the study of finite density QCD at low values of the
gauge coupling. The situation may of course be quite different
depending on the temporal extent $L_t$ of the lattice, on 
the number of flavors, of colors and on the lattice discretization
(staggered or Wilson fermions) adopted. We plan to make a more extensive 
study of this problem in the future.

\begin{table}
\begin{center}
\begin{tabular}{|c|c|c|}
\hline $L_s$ & $\beta_c$ & $\mu_c$ \\
\hline  $8^3\times 6$ &  $1.584(2) $  & $0$ \\
\hline $12^3\times 6$ &  $1.587(2) $  & $0$ \\
\hline $16^3\times 6$ &  $1.582(2) $  & $0$ \\
\hline $16^3\times 6$ &  $1.568(2) $  & $0.15$    \\
\hline  $8^3\times 6$ &  $1.55$    &  $0.222(10) $ \\   
\hline $16^3\times 6$ &  $1.55$    &  $0.215(10) $ \\
\hline  $8^3\times 6$ &  $1.5$     &  $0.325(10) $ \\
\hline $12^3\times 6$ &  $1.5$     &  $0.349(15)$ \\
\hline $16^3\times 6$ &  $1.5$     &  $0.342(10) $ \\
\hline 
\hline   & $\beta_{Sc}$ & $\mu_{Sc}$ \\
\hline  $8^3\times 6$ &  $1.55$    &  $0.678(20)$ \\
\hline  $8^3\times 6$ &  $1.675$   &  $0.793(10)$  \\
\hline $16^3\times 6$ &  $1.675$   &  $0.789(22)$ \\
\hline
\end{tabular}
\end{center}
\caption{\label{couplings}
Collection of pseudocritical couplings as determined
from our numerical data. Physical critical couplings have
been determined through the chiral susceptibility, while 
the unphysical saturation transitions have been located
by means of the plaquette susceptibility. 
}
\end{table}

\subsection{Did we catch the physics of the low temperature 
region of the phase diagram?}
\label{lowtemp}

One of our starting questions was about the fate of confinement
at high densities and low temperatures, since that could help 
understanding the nature of compact astrophysical objects.
In Ref.~\cite{Hands06} the hypothesis has been made, based on
the analysis of the Polyakov loop, that at $T \sim 0$ deconfinement could occur
at a critical density following and well separated from the onset
of a bosonic superfluid phase. It is natural to ask whether our 
present results can be of any relevance regarding this specific
issue, \ie how close we have got to the low temperature
region of the QCD phase diagram.

Since we have not included an explicit diquark source term in our
model, we cannot obtain direct information about that observable;
however we shall try to sketch a qualitative picture based 
on the distribution of the eigenvalues of the fermionic matrix.
At zero density that can be written as
$M = a m\, {\rm Id} + D$ where $D$ is antihermitian, hence it has purely
imaginary eigenvalues, therefore the eigenvalues of $M$ lie on a
segment in the complex plane orthogonal to the real axis. 

As a real chemical potential is switched on, $D$ ceases to be 
antihermitean and the eigenvalues get scattered in the whole complex
plane: that is evident in the first inset of Fig.~\ref{eigendistr},
where we show the distribution of eigenvalues on a typical configuration
obtained at $\beta = 1.55$ and a small chemical potential, 
$\hat \mu = 0.10$. The eigenvalues occupy a narrow vertical band 
and the finite density of eigenvalues
in correspondence of the real axis is strictly linked to the presence
of chiral symmetry breaking (Banks-Casher relation~\cite{banks})
 The width of the distribution on the real 
axis grows as $\hat \mu$ increases, roughly proportionally to  
$\hat \mu^2$, till the distribution
touches the imaginary axis: at this point the chiral condensate
is expected to rotate into a diquark condensate
(see for instance Ref.~\cite{splitt} for a review):
as it is clear from the second inset in Fig.~\ref{eigendistr},
at $\beta = 1.55$ this happens roughly at $\hat \mu \sim 0.3$,
a value which actually turns out to be almost independent of the gauge
coupling in the range of $\beta$ values explored in our simulations
and is in agreement with the values found for diquark condensation
in similar works using the same quark mass~\cite{Hands99,bittner}:
we show as an example in Fig.~\ref{realdens} the eigenvalue distribution 
projected onto the real axis for three values of $\hat \mu$
at $\beta = 1.45$.

\begin{figure}[b!]
\includegraphics*[angle=-90,width=0.9\columnwidth]{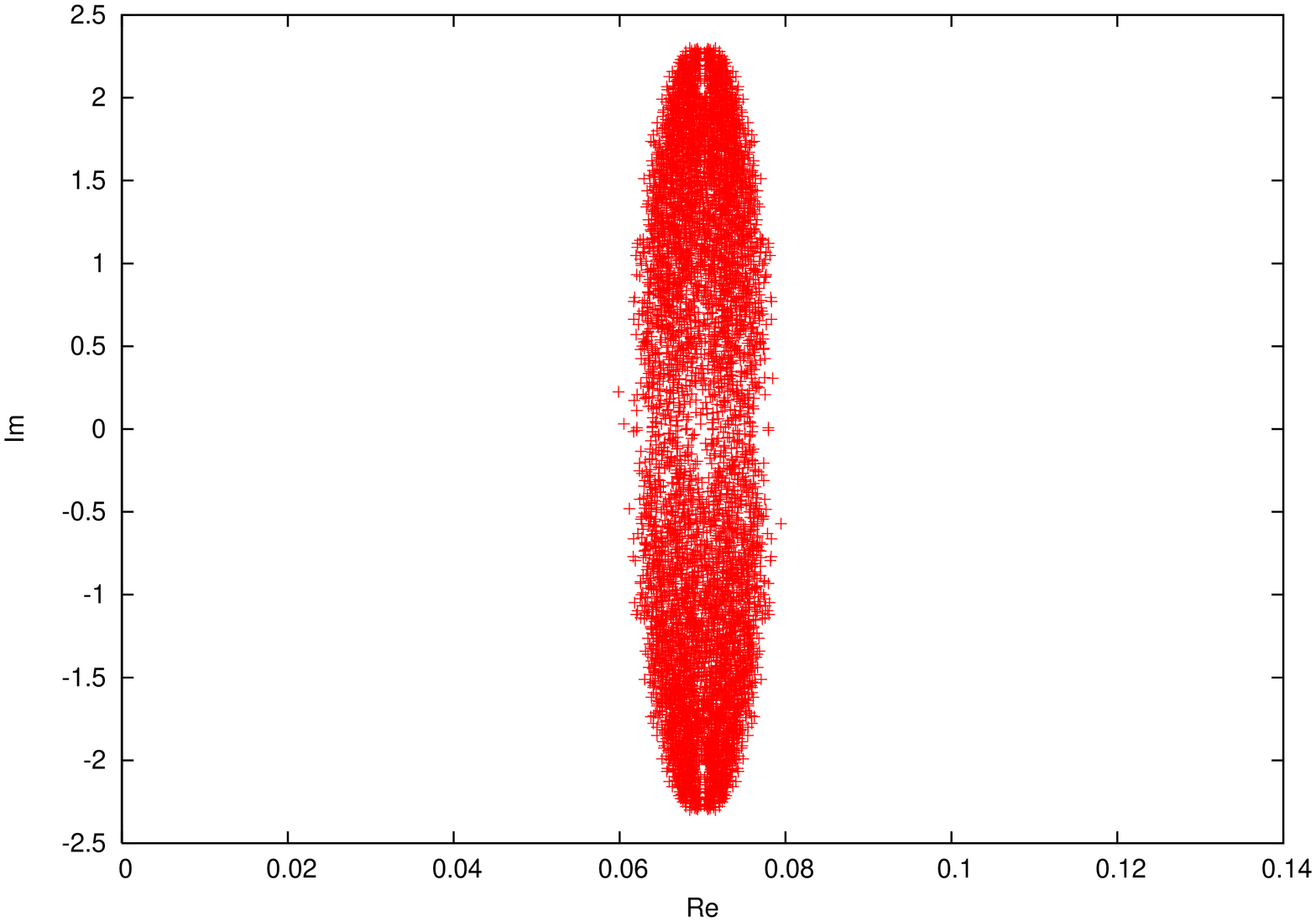}\\
\includegraphics*[angle=-90,width=0.9\columnwidth]{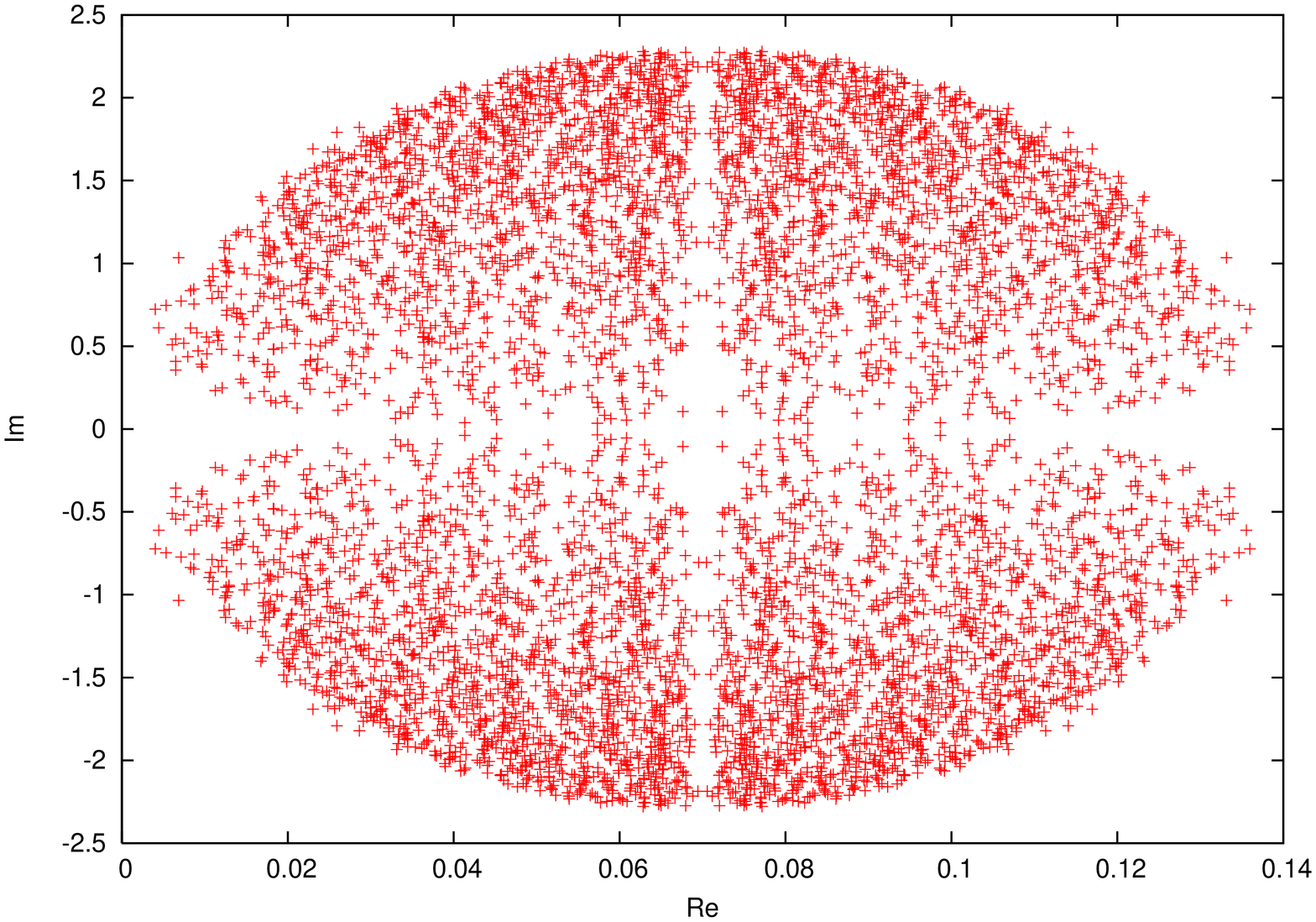}\\
\includegraphics*[angle=-90,width=0.9\columnwidth]{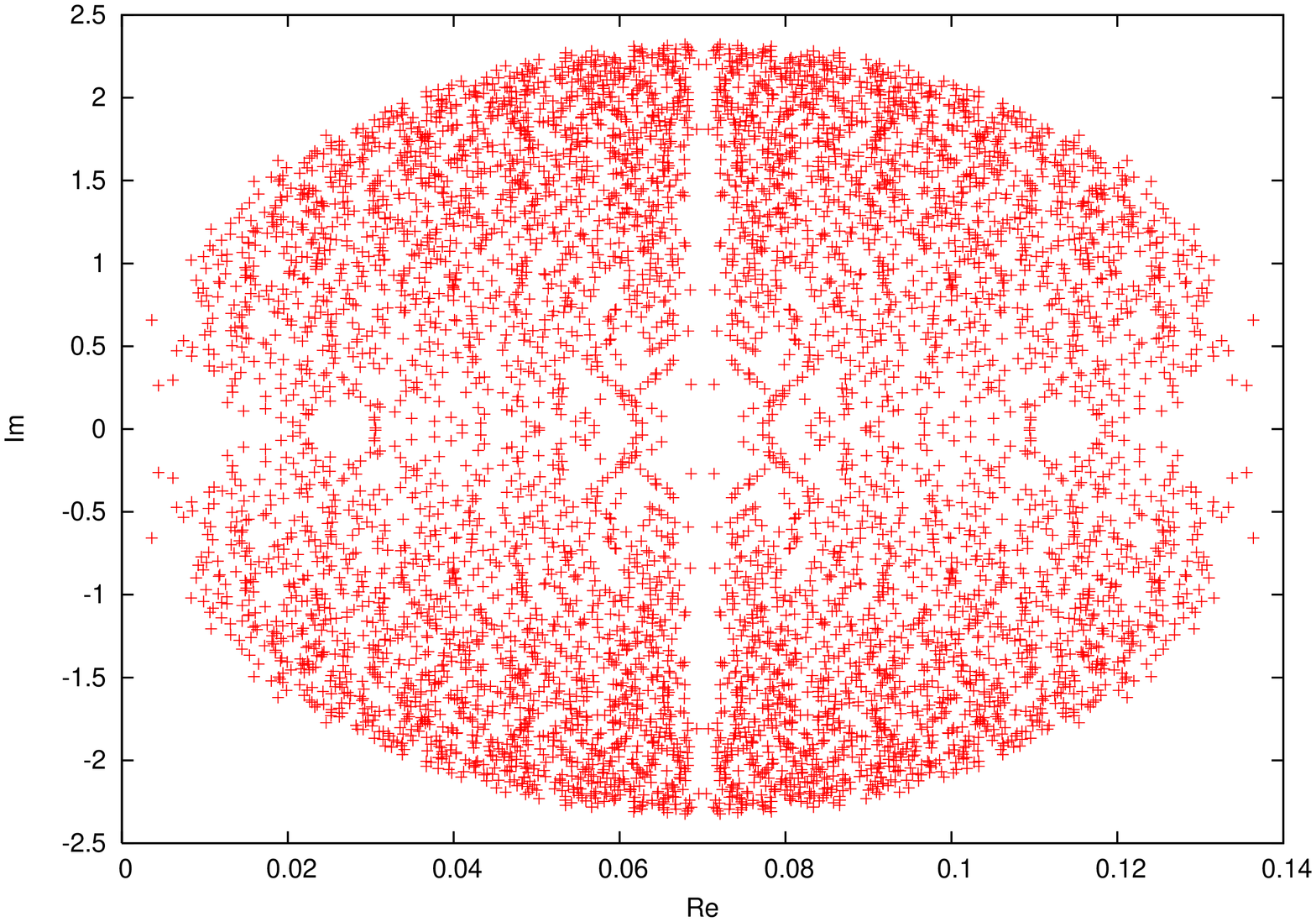}\\
\vspace{-0.cm}
\caption{Distribution of eigenvalues on typical configurations 
of the $8^3 \times 6$ lattice obtained respectively 
at ($\beta = 1.55$, $\hat\mu = 0.10$), 
($\beta = 1.55$, $\hat\mu = 0.30$) and  
($\beta = 1.45$, $\hat\mu = 0.30$).}
\label{eigendistr} 
\vspace{-0.cm}
\end{figure}

\begin{figure}[b!]
  \includegraphics*[angle=-90,width=1.0\columnwidth]{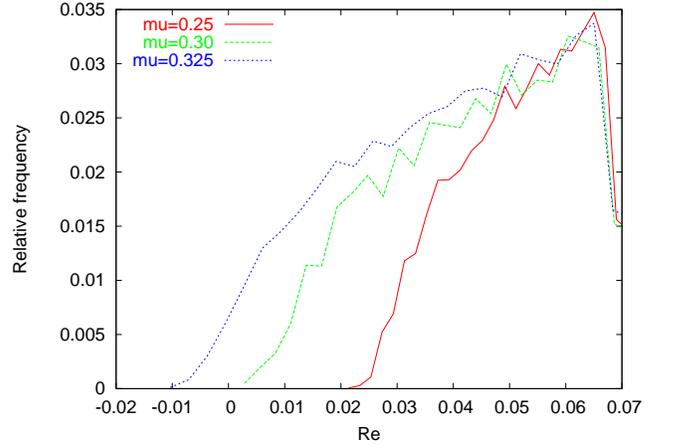}
\vspace{-0.cm}
\caption{
Distribution of eigenvalues projected onto the real axis for various
$\hat\mu$ on a $8^3 \times 6$ lattice at $\beta = 1.45$.
}
\label{realdens} 
\vspace{-0.cm}
\end{figure}

However if we are in the high temperature
region, \ie slightly below $T_c (\mu = 0)$, 
chiral symmetry will be restored quite soon as $\hat\mu$ is increased
because of the transition to the 
Quark-Gluon Plasma. Therefore  
there will be actually no chiral condensate 
to be rotated into a diquark condensate at the point where the
distribution touches the imaginary axis. Indeed we see from 
the second inset in Fig.~\ref{eigendistr} that the region 
around the real axis is quite depleted of eigenvalues for 
$\beta = 1.55$ at $\hat\mu = 0.3$.
We can easily understand this in terms of the chiral line we
have drawn in Fig.~\ref{diagr}: chiral symmetry gets restored already
below $\hat \mu = 0.3$ at $\beta = 1.55$.

Following this line of reasoning, the region relevant for low
temperature physics on our lattices with $L_t = 6$ should be that
below $\beta \sim 1.5$, where our fitted 
(pseudo)critical line passes beyond $\hat \mu \sim 0.3$.
In this region one could for instance observe, among other
different possibilities, two different
transitions, the first corresponding to the onset of diquark
condensation, the second roughly being the continuation
of the line in Fig.~\ref{diagr}, thus corresponding to deconfinement:
this is indeed the scenario suggested by Ref.~\cite{Hands06}.

\begin{figure}[b!]
  \includegraphics*[angle=-90,width=1.0\columnwidth]{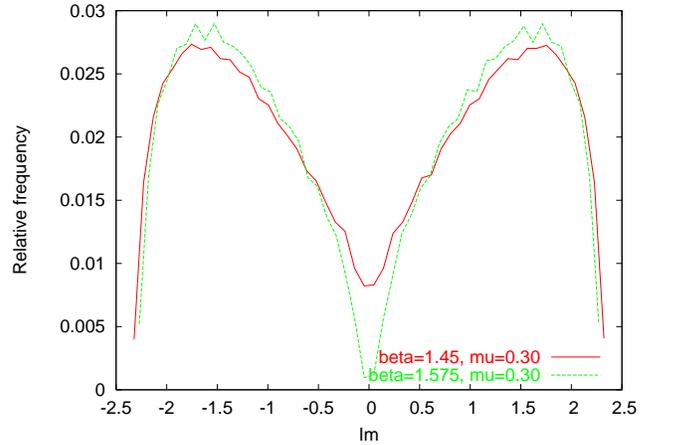}
\vspace{-0.cm}
\caption{
Distribution of eigenvalues projected onto the imaginary axis
for various values of $\beta$  
on a $8^3 \times 6$ lattice at $\hat\mu = 0.3$.
}
\label{imagdens} 
\vspace{-0.cm}
\end{figure}

We have therefore performed numerical simulations on a
$8^3 \times 6$ lattice at $\beta = 1.45$. In this case
a finite density of eigenvalues around the real axis 
is still present at $\hat \mu \sim 0.3$, as can be better appreciated
in Fig.~\ref{imagdens}, where we plot the distribution projected
onto the imaginary axis at $\beta = 1.45$ and $\hat\mu = 0.3$,
compared to that obtained at higher temperatures.

In Fig.~\ref{fig:suscet145} we report the chiral susceptibility,
compared to that measured on the same lattices at different gauge 
couplings. We notice that the peak is strongly reduced
and its position is not much different from what obtained at 
$\beta = 1.5$ and in clear disagreement with what expected from the 
continuation of the chiral line in Fig.~\ref{diagr}.
The first peak could indeed correspond to the onset of a 
bosonic superfluid phase. Nothing seems to happen thereafter.

In Fig.~\ref{fig:rho08} we report instead data obtained for 
the susceptibility $\rho_D$ of the disorder parameter.
In this case the negative peak has almost completely disappeared and
a very small peak at $\hat \mu \sim 0.3$ is followed by a region 
$\hat \mu \geq 0.4$ where
$\rho_D$ clearly changes its sign: on the basis of what we have discussed
in Section~\ref{saturation} and comparing this behaviour with that
observed at $\beta = 1.55$, a possible interpretation is that  
of an early onset of saturation effects in this case, preventing 
the observation of any further physical transition.
We expected saturation effects to obscure the physical
transition at $\beta \sim 1.4$, but we are not surprised that the
situation may be worse. 

This conclusion is supported by looking at the behaviour of the
Polyakov loop (see Fig.~\ref{fig:poly08}), in this case
saturation effects are signalled by an inversion
in the growth of $\langle L \rangle$ as a function of $\hat\mu$.

We conclude therefore that we are not be able to clarify the 
onset of deconfinement at $T \sim 0$, at least on the present
lattice size. We could of course further decrease the temperature
without decreasing $\beta$ by going to larger values of $L_s$.
However that would imply a numerical effort which is not affordable
with our present algorithmic and computational resources.

\begin{figure}[b!]
\includegraphics*[width=1.0\columnwidth]{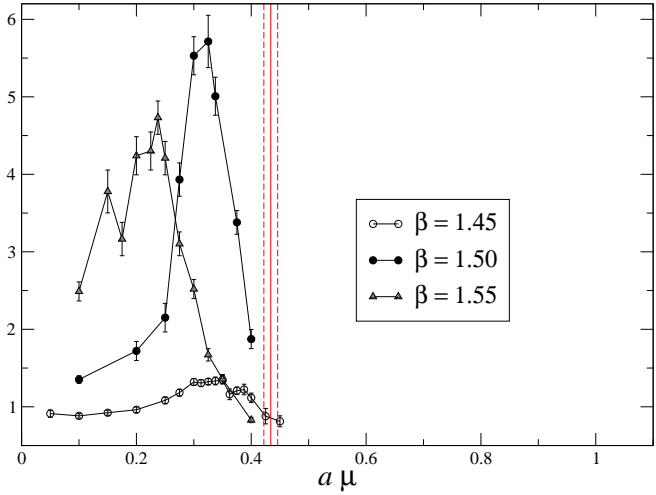}
\vspace{-0.cm}
\caption{
Chiral susceptibilities on the $8^3\times 6$ lattice at various values
of $\beta$.
}
\label{fig:suscet145} 
\vspace{-0.cm}
\end{figure}

\begin{figure}[b!]
\includegraphics*[width=1.0\columnwidth]{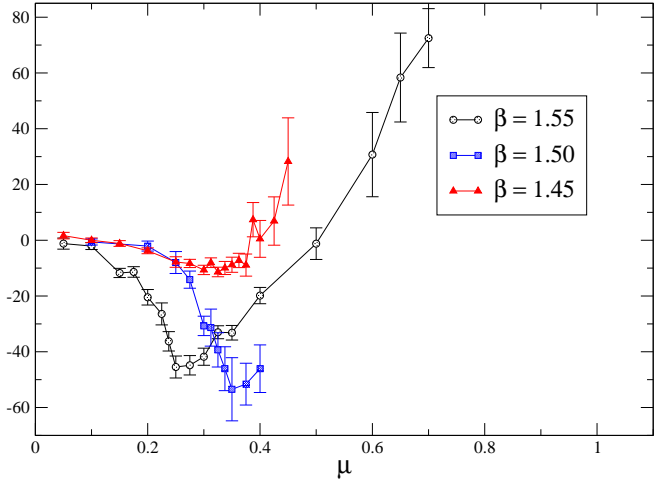}
\vspace{-0.cm}
\caption{
$\rho_D$ at various values of $\beta$ on a $8^3\times 6$ lattice.
}
\label{fig:rho08} 
\vspace{-0.cm}
\end{figure}

\begin{figure}[b!]
\includegraphics*[width=1.0\columnwidth]{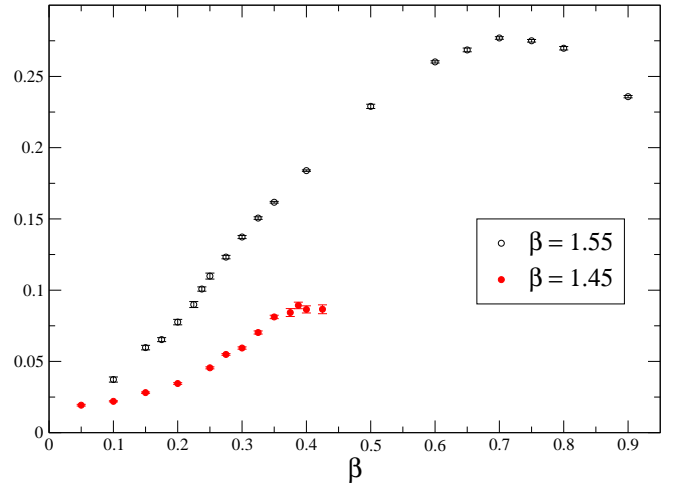}
\vspace{-0.cm}
\caption{
Polyakov loop at various values of $\beta$ on a $8^3\times 6$ lattice.
}
\label{fig:poly08} 
\vspace{-0.cm}
\end{figure}

\section{Conclusions} 
\label{conclusions}

We have investigated the phase diagram of two-color QCD
at finite temperature and density by means of a disorder 
parameter for color confinement detecting dual superconductivity 
of the QCD vacuum.

We have obtained evidence for deconfinement induced by a finite
density of baryonic matter. Moreover the transition line
corresponding to the disappearance of dual superconductivity
(deconfinement) appears to coincide, in the range of temperature
explored ($ 0.4\, T_c < T < T_c$, where $T_c$ is the critical temperature
at zero density), with that corresponding to 
chiral symmetry restoration, as it happens in the zero density 
case. We have also shown that the susceptibilities of the disorder
parameter can be used in order to compute the slope of the critical
line in the $\beta - \hat\mu$ plane, obtaining consistent results.

We have investigated in some detail the unphysical transition 
corresponding to the onset of saturation and shown that it moves
at lower values of $\hat\mu$ as $\beta$ is decreased
with a possible intersection with the physical transition line, 
thus giving a general warning about the possible effects
of saturation on the study of finite density QCD at
strong values of the gauge coupling. This phenomenon of course
may be quite different depending on the fermion discretization,
on the number of flavors and on other parameters of
the system ($L_t$, quark masses); for this reason we plan 
to make a more systematic study in the future.

We have also verified that in our case saturation actually
prevents us from obtaining results relevant for the $T \sim 0$
region of the phase diagram. For this reason we plan to extend
our study in the future by adopting different lattice sizes
and/or fermion discretizations. 
Results relevant for the high temperature region of 
real QCD could also be obtained within the imaginary 
chemical potential approach.

\begin{acknowledgments}
We thank Ph. de Forcrand, A. Di Giacomo, M.P. Lombardo and B. Lucini
for useful discussions. Numerical simulations have been run on a PC
farm at INFN-Genova and on the INFN apeNEXT facility in Rome.
Simulations on the smaller lattices have been performed by means of
a numerical code obtained by adapting the publicly available MILC code.
\end{acknowledgments}

\end{document}